\documentclass[sigconf]{acmart}
\AtBeginDocument{%
  }

\usepackage{graphicx}
\usepackage{xcolor}
\usepackage{amsmath}
\usepackage{fontawesome}
\usepackage{balance}

\usepackage{subcaption}
\usepackage{svg}

\copyrightyear{2026}
\acmYear{2026}
\setcopyright{cc}
\setcctype{by-nc-nd}
\acmConference[SIGIR '26]{Proceedings of the 49th International ACM SIGIR Conference on Research and Development in Information Retrieval}{July 20--24, 2026}{Melbourne, VIC, Australia}
\acmBooktitle{Proceedings of the 49th International ACM SIGIR Conference on Research and Development in Information Retrieval (SIGIR '26), July 20--24, 2026, Melbourne, VIC, Australia}
\acmDOI{10.1145/3805712.3808569}
\acmISBN{979-8-4007-2599-9/2026/07}

\begin{document}

\title{Lost in the Evidence? Reproducing Document Position and Context Size Effects in RAG}

\author{Jorge Gabín}
\email{jorge.gabin@udc.es}
\orcid{0000-0002-5494-0765}
\affiliation{%
  \institution{IRLab, CITIC, Universidade da Coruña}
  \city{A Coruña}
  \country{Spain}
}
\affiliation{
    \institution{Linknovate Science}
    \city{Santiago de Compostela}
    \country{Spain}
}

\author{Anxo Perez}
\email{anxo.pvila@udc.es}
\orcid{0000-0002-0480-006X}
\affiliation{%
  \institution{IRLab, CITIC, Universidade da Coruña}
  \city{A Coruña}
  \country{Spain}
}

\author{Javier Parapar}
\email{javier.parapar@udc.es}
\orcid{0000-0002-5997-8252}
\affiliation{%
  \institution{IRLab, CITIC, Universidade da Coruña}
  \city{A Coruña}
  \country{Spain}
}


\begin{abstract}
Retrieval-Augmented Generation (RAG) systems rely on retrieved documents being concatenated into a model’s input context, making both document ordering and context size critical yet controversial design choices. Prior work reports position-based effects such as \textit{lost in the middle} and related long-context phenomena. However, empirical findings remain inconsistent and hard to reproduce across models, datasets, and evaluation protocols. In this paper, we present a systematic reproducibility study that revisits these claims and examines how they evolve with contemporary LLMs under a controlled evaluation framework. We first show that topic sampling is a major source of variance: small topic sets can mask or exaggerate ordering effects. Based on repeated subset sampling across multiple topic budgets, we provide a practical calibration procedure that identifies topic counts yielding stable trends at feasible cost. Using these fixed topic sets, we then reproduce and extend results on position sensitivity, re-evaluating \textit{lost in the middle} and positional biases in modern LLMs. Then, we also study a more realistic RAG scenario in which relevance is mediated by a retriever rather than oracle access to ground-truth documents. In this setting, we re-examine a recent industry study and identify discrepancies to evaluation choices such as limited topic coverage and reliance on LLM-based judges. Finally,  we conduct an analysis of how retrieval order and context size affect downstream LLM performance under imperfect retrieval. Our results demonstrate that both factors interact strongly with retrieval quality and model choice, and that conclusions drawn from idealised setups do not always transfer to real-world RAG pipelines. We release all code and configurations to support reproducibility and future work on robust RAG evaluation.
\end{abstract}




\maketitle

\section{Introduction}

Large Language Models (LLMs) have become fundamental components of many industrial and research applications~\cite{microsoft2025rag}. Among them, Retrieval-Augmented Generation (RAG) is a key paradigm for enabling LLMs to answer questions grounded in external knowledge~\cite{Lewis-et-al-2020,karpukhin-etal-2020-dense}. By retrieving relevant passages and injecting them into the model’s input context, RAG systems effectively extend factual coverage and adaptability. However, despite widespread adoption, current implementations often fail to exploit their full potential. In particular, critical aspects of RAG systems, such as the optimal number of passages to retrieve (context size) and the order in which they should be presented, remain uncertain~\cite{Yu2024InDO,cao-bloomberg-2025,liu-etal-2024-lost-in-the-middle,DBLP:conf/ecir/TianGM25,DBLP:conf/ecir/HutterRMK25}.

Prior studies have analysed the context size and passage order effect on RAG systems, reaching conflicting conclusions: some report strong gains from placing the most relevant passages first~\cite{liu-etal-2024-lost-in-the-middle}, or from preserving intra-document order~\cite{Yu2024InDO}. Other authors claim little to no sensitivity to ordering or retrieval depth, with trends varying by model and dataset~\cite{cao-bloomberg-2025,liu-etal-2024-lost-in-the-middle,DBLP:conf/ecir/TianGM25,DBLP:conf/ecir/HutterRMK25}. These discrepancies are difficult to interpret because prior work often differs in the topic sets used in their experiments, the evaluation protocol (automatic metrics vs. LLM-based judges), and the LLMs and retrieval pipelines under study. Thus, practitioners still face uncertainty regarding crucial configuration choices. Given the increasing reliance of industry on these systems~\cite{lithgow2025assessing,xu2024generateive}, a clear understanding of how these design factors influence results is of paramount importance.

% Motivated by these inconsistent findings, our goal is to provide reliable and reproducible evidence on how context size and passage ordering affect RAG performance. Specifically, we focus on two widely used question answering benchmarks: Natural Questions (NQ)~\cite{kwiatkowski-etal-2019-natural}, which is predominantly single-hop, and HotpotQA~\cite{yang-etal-2018-hotpotqa}, which requires multi-hop evidence aggregation. Our starting point is a methodological observation: conducting experiments on full topic sets is frequently impractical, especially when comparing multiple LLMs, yet small topic samples can yield unstable conclusions. We therefore introduce a calibration procedure for selecting an \emph{adequate topic budget}. We evaluate multiple subset sizes and, for each size, repeatedly sample random topic subsets to measure how sensitive ordering and context size trends are to topic choice. This approach enables us to estimate the minimum number of topics required to obtain stable and reliable performance measurements, allowing subsequent experiments with other models or configurations to be conducted efficiently while maintaining reproducibility.

Motivated by these inconsistent findings, our goal is to provide reliable and reproducible evidence on how context size and passage ordering affect RAG performance. Specifically, we focus on two widely used question answering benchmarks: Natural Questions (NQ)~\cite{kwiatkowski-etal-2019-natural}, which is predominantly single-hop, and HotpotQA~\cite{yang-etal-2018-hotpotqa}, which requires multi-hop evidence aggregation. Our starting point is a methodological observation: conducting experiments on full topic sets is frequently impractical, especially when comparing multiple LLMs. However, an arbitrary selection of small topic samples can yield unstable conclusions.

We therefore introduce a calibration procedure for selecting an \emph{adequate topic budget} based on the stability of performance trends. We evaluate multiple subset sizes $n$ and, for each size, repeatedly sample random topic subsets to measure how sensitive ordering and context size trends are to the specific topics chosen. To ensure reliability, we identify the threshold where the variance of the \emph{$\Delta$F1} between sorting strategies is sufficiently low to prevent ``\textit{zero-crossings}'', instances where the relative ranking of two methods flips due to sampling noise. This approach enables us to obtain stable performance, allowing subsequent experiments with other models or configurations to be conducted reliably and reproducibly.

Using this controlled evaluation framework, we then revisit prominent claims in the literature. We reproduce and extend long-context position-effect studies~\cite{cao-bloomberg-2025,liu-etal-2024-lost-in-the-middle} under modern LLMs, and we re-examine recent robustness claims in realistic RAG scenarios where relevance is mediated by imperfect retrieval rather than oracle access to gold documents~\cite{Yu2024InDO}. Beyond reproductions, we also assess the impact of the retrieval phase itself, testing both standard ranking models and oracle rankers, as well as more advanced retrieval strategies, to determine how crucial retrieval quality is for overall performance. Additionally, we examine how different LLM architectures and sizes behave under these varying setups. By systematically exploring these factors, we aim to clarify under which configurations RAG systems perform reliably and when their outputs are sensitive to design choices.

Our contributions are threefold: (1) a dataset-specific calibration method for choosing topic sizes that yield stable conclusions when studying ordering and context-size effects; (2) a systematic reproducibility study that revisits positional-bias and RAG robustness claims under this controlled evaluation framework and contemporary LLMs. (3) Finally, an extensive analysis of how retrieval quality, ordering, context size, dataset characteristics, and model family/scale influence Question Answering (QA) performance. We release all code and configurations to support future reproducible research\footnote{\faicon{github} \href{https://github.com/IRLab-UDC/Lost-in-the-Evidence-in-RAG-Virtual-Appendix}{https://github.com/IRLab-UDC/Lost-in-the-Evidence-in-RAG-Virtual-Appendix}}.

\section{Related Work}

Retrieval-Augmented Generation combines a retriever with a generator to answer knowledge-intensive queries by grounding outputs in external evidence. Early and influential architectures include REALM~\cite{guu-REALM-2020}, and the neural RAG framework of Lewis et al.~\cite{Lewis-et-al-2020}. On the retrieval side, sparse baselines such as BM25 and dense pipelines that pair dual-encoder retrieval with cross-encoder reranking remain standard in practice~\cite{bm25-robertson,karpukhin-etal-2020-dense,Nogueira2019PassageRW}. Recent surveys synthesise common RAG components (retriever, composer, generator), pipeline variants (e.g., joint vs.\ fixed retrievers), and evaluation practices, highlighting open issues around robustness, context construction, and measurement choices~\cite{hau-rag-survey-2024,gao2023retrieval}.

A growing body of evidence shows that LLMs are sensitive to where evidence appears in long contexts. The \textit{lost in the middle} phenomenon shows that moving the same answer-bearing content to different positions can change QA accuracy, often peaking near the beginning or end and degrading in the middle~\cite{liu-etal-2024-lost-in-the-middle}. Follow-up work argues that position effects interact with how passages are composed, including whether chunks preserve intra-document order versus being concatenated purely by relevance, and reports non-monotonic behaviour as more chunks are added~\cite{Yu2024InDO}. Complementary to purely positional studies, a growing line of RAG robustness work evaluates end-to-end pipelines under systematic perturbations to retrieved evidence, including retrieval depth ($k$) and permutations of the retrieved list~\cite{cao-bloomberg-2025}. These studies often find that average trends can appear stable, yet robustness is imperfect and can hide substantial instance-level trade-offs when order or $k$ changes~\cite{cao-bloomberg-2025}. Another closely related direction studies how retrieval quality and noise shape generation. Because retrieval is imperfect, retrieved contexts naturally contain irrelevant or weakly relevant passages, which can distract the generator and degrade answer correctness, especially when such passages are prominent or placed in salient positions~\cite{amiraz-etal-2025-distracting}. More broadly, recent work on RAG evaluation emphasises that conclusions can depend strongly on choices such as dataset formatting, chunking/composition, and scoring methodology,  and that resource constraints often lead to evaluations on reduced subsets whose representativeness is rarely validated~\cite{hau-rag-survey-2024}.

Taken together, prior work suggests that retrieval depth and ordering interact with model biases and retrieval quality, and that conclusions may depend on dataset/sample size, score distributions, and prompt format. We study all these factors in our reproducibility study. We differ from past reports by (i) reproducing recent claims about weak order/depth effects under matched settings, (ii) isolating sources of variance to obtain stable results, and (iii) extending the analysis across ordering schemes, context sizes, retrieval quality (BM25 vs.\ dense reranking vs.\ oracle contexts), and model families/sizes.

\section{Towards Stable Evaluation of Context Ordering and Size}
\label{sec:towards-stable-evaluation}

As mentioned, prior work on context size ($k$) and evidence ordering in LLM-based QA reports inconsistent trends,  and it is often unclear whether disagreements reflect genuine model behavior or instability in evaluation protocols~\cite{liu-etal-2024-lost-in-the-middle,Yu2024InDO,cao-bloomberg-2025,DBLP:conf/ecir/HutterRMK25}. Because running full-topic evaluations is often impractical (especially when comparing multiple LLMs or using expensive judging schemes) many papers rely on relatively small subsets of topics~\cite{liu-etal-2024-lost-in-the-middle,DBLP:conf/ecir/HutterRMK25,cao-bloomberg-2025}. However, if topic sampling itself introduces large variance, then conclusions about order and the number of retrieved documents $k$ can be unstable and difficult to reproduce. In this section, we therefore introduce a simple calibration procedure to determine an \emph{adequate topic budget} for studying context-ordering and context-size effects. The goal is to identify the smallest number of topics that yield stable conclusions while keeping computational cost manageable. We then adopt the resulting topic counts as a controlled evaluation setting for all subsequent reproducibility experiments in the paper.

\paragraph{\textbf{Setup.}} We focus on two widely used QA benchmarks: NQ~\cite{kwiatkowski-etal-2019-natural}, which primarily consists of single-hop queries where the answer is typically contained within a single passage, and HotpotQA~\cite{yang-etal-2018-hotpotqa}, a more complex dataset requiring multi-hop reasoning to aggregate evidence across multiple distinct passages. We employ LLaMA-3.1:8B as our experimental model, selected for its status as a modern, widely adopted standard in recent literature, offering a balance of high performance and manageable medium-scale parameter size. For each query, we construct a context by retrieving the top-$k$ passages and concatenate them into the prompt. Similar to prior works, we study three standard ordering schemes applied to a fixed top-$k$ set: \emph{standard} (descending retrieval score), \emph{reverse} (ascending score), and \emph{random} (uniform permutation). Our analysis uses F1-token (hereafter F1) as the primary metric, since it is the official benchmark metric and does not depend on a particular judge model.

\subsection{Investigating Sources of Variability}

\begin{figure}[t]
    \centering
    \begin{subfigure}[b]{0.49\textwidth}
        \centering
        \includegraphics[width=\textwidth]{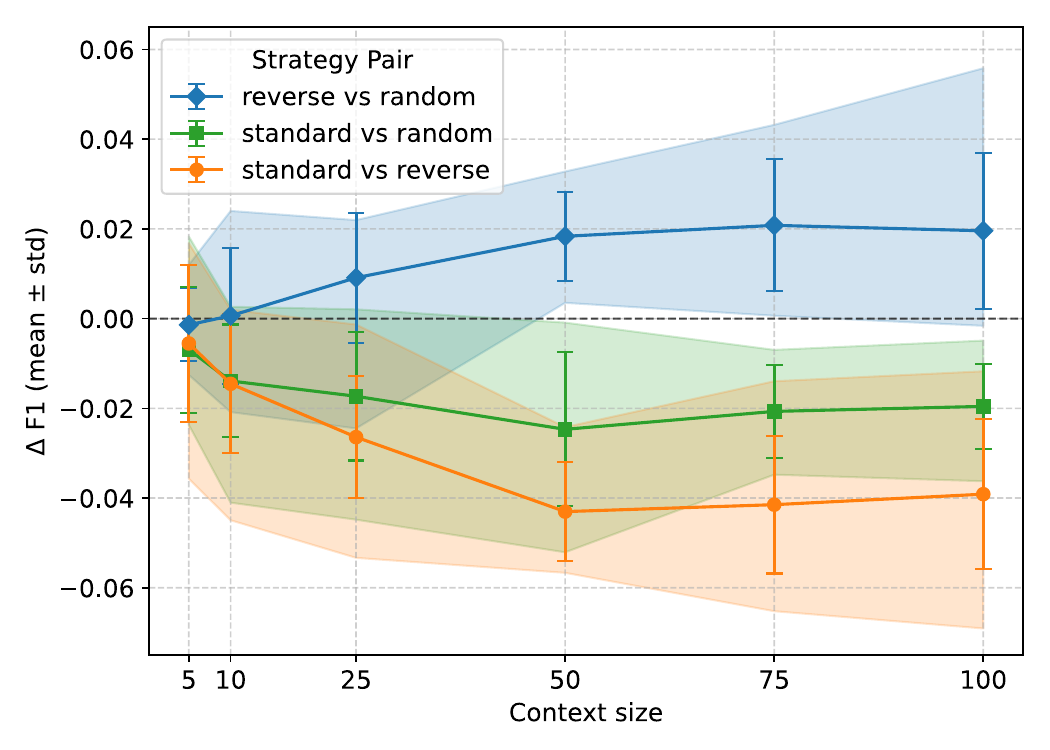}
        \caption{HotpotQA (500 topics)}
        \label{fig:rq0_1_hotpot}
    \end{subfigure}
    \hfill
    \begin{subfigure}[b]{0.49\textwidth}
        \centering
        \includegraphics[width=\textwidth]{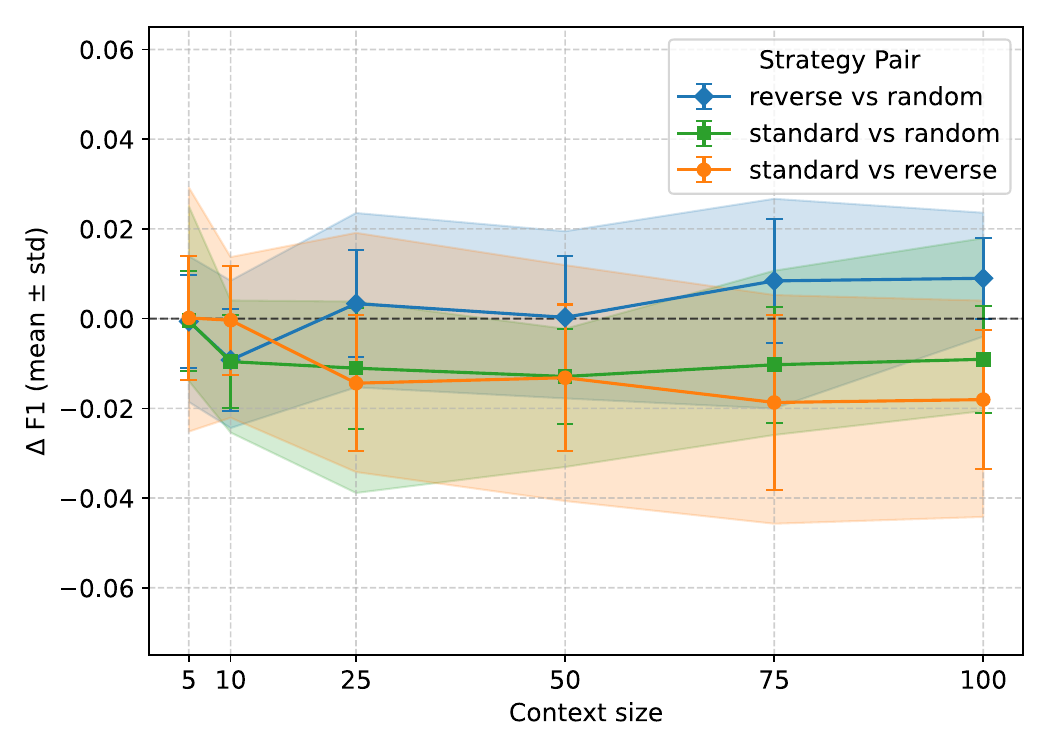}
        \caption{NQ (500 topics)}
        \label{fig:rq0_1_nq}
    \end{subfigure}
    \caption{Performance variability across 10 random subsets of 500 topics for HotpotQA and NQ. The figure shows the $\Delta$F1 between ordering strategies at different context sizes. Dots indicate the mean $\Delta$F1, error bars represent the standard deviation, and shaded areas denote the minimum/maximum values across subsets.}
    \label{fig:rq0-1}

\end{figure}

We first quantify how sensitive ordering conclusions are to the particular set of evaluation topics. For a fixed subset size $n$, we repeatedly sample topic subsets using different random seeds. For each subset, we evaluate all context sizes and ordering schemes and compute $\Delta$F1, which is the pairwise F1 difference between strategies (e.g., reverse–standard, reverse–random, standard–random).

Figures~\ref{fig:rq0_1_hotpot} and \ref{fig:rq0_1_nq} summarise this analysis over 10 random subsets with $n\!=\!500$ topics per dataset. We use 500 topics because it is a common evaluation budget in recent RAG work (e.g., \cite{cao-bloomberg-2025}), and it lets us test whether a seemingly reasonable topic size can still yield unstable conclusions. In the plots, dots show the mean \emph{$\Delta$F1} across subsets, error bars denote the standard deviation, and the shaded region indicates the min–max range. The x-axis is context size and the y-axis is \emph{$\Delta$F1}. With $n\!=\!500$, we can see the variability is large enough that the apparent winner among ordering strategies can change across samples, particularly at smaller context sizes.

This implies that conclusions about whether ``order matters'' can depend heavily on which topics happen to be included in the evaluation, providing a plausible explanation for discrepancies across prior reports that use different subsets and protocols. This leads us to our next point, where we study the adequate number of topics and queries to ensure that observed trends are robust and representative.

\begin{figure}[t]
    \centering
    \begin{subfigure}[b]{0.49\textwidth}
        \centering
        \includegraphics[width=\textwidth]{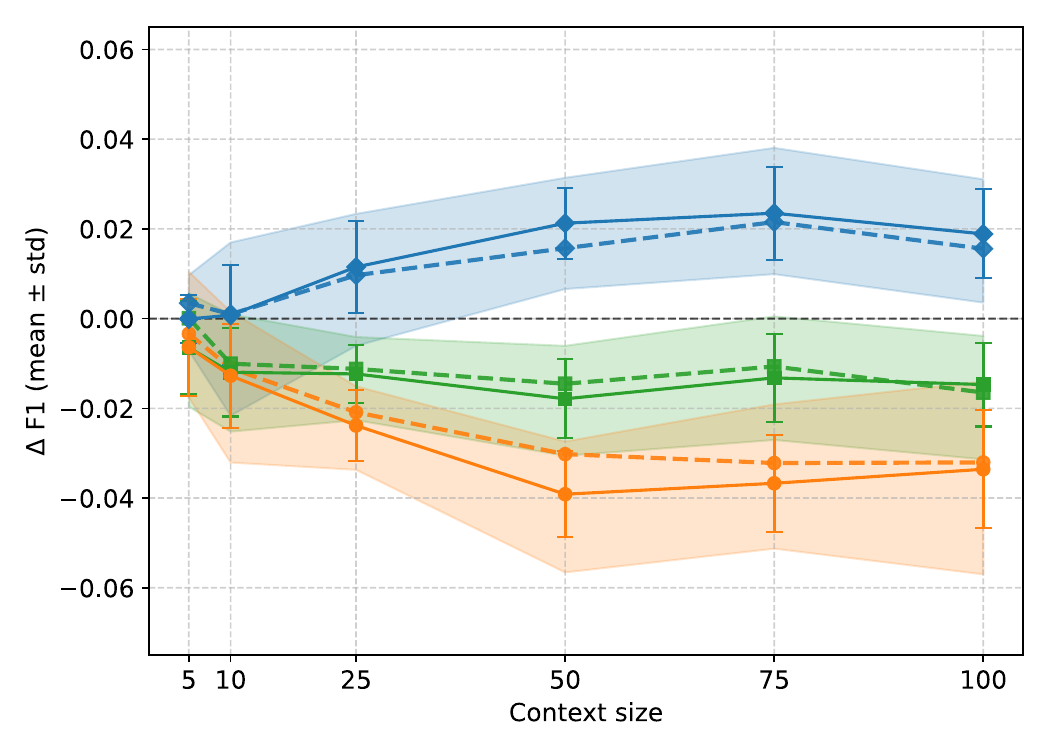}
        \caption{HotpotQA (1000 topics)}
        \label{fig:rq0_2_hotpot}
    \end{subfigure}
    \hfill
    \begin{subfigure}[b]{0.49\textwidth}
        \centering
        \includegraphics[width=\textwidth]{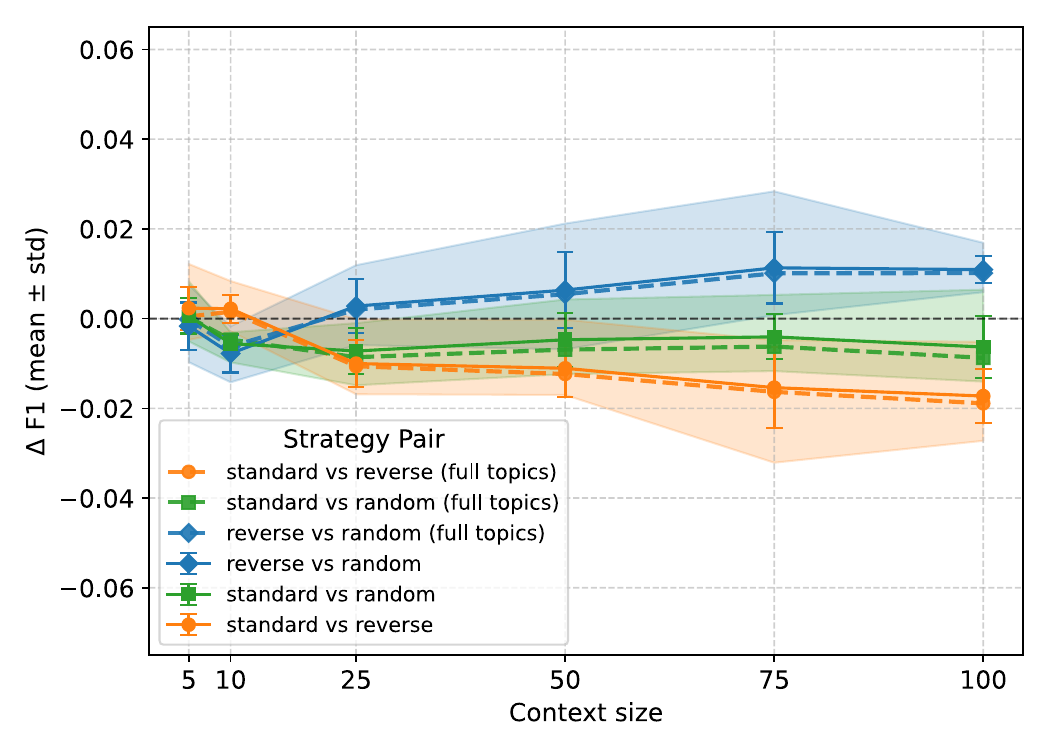}
        \caption{NQ (2000 topics)}
        \label{fig:rq0_2nq}
    \end{subfigure}
    \caption{Performance variability across 10 random subsets of 1000 and 2000 topics for HotpotQA and NQ, respectively. In addition to Figure~\ref{fig:rq0-1}, the dashed line represents the $\Delta$F1 using the full topics.}
    \label{fig:rq0_2}
\end{figure}

\subsection{Determining an Adequate Number of Topics}

We next translate our variability diagnosis into a practical calibration rule: what is the minimum number of topics required to ensure that the comparative rankings between sorting strategies are stable? We aim to identify the smallest topic set size where evaluation findings become representative. 

We begin by establishing a ground truth reference through a single evaluation over the full topic set for each dataset, which serves as the upper bound for stability. Next, we systematically vary the sample size ($n \in \{500, 1000, 2000, 3000, 4000, 5000\}$) and measure the variance in $\Delta$F1 performance between ordering schemes across random seeds. Crucially, we focus on minimising the frequency with which these performance deltas cross the zero line. A zero-crossing implies that the ``winning'' strategy flips purely due to sampling noise, indicating that the sample size is insufficient to discern a reliable difference. Frequent oscillations around zero suggest that the relative ranking of methods is unstable. Conversely, a reliable sample size yields a $\Delta$F1 that remains consistently positive or negative, ensuring that the ordering of methods is deterministic rather than an artefact of the specific subset chosen.

Our experiments reveal that increasing the number of topics progressively decreases these oscillations, eventually reducing the variance enough that the performance deltas no longer intersect the zero line for noticeable differences. This transition marks the practical threshold for reliable evaluation. Figure~\ref{fig:rq0_2} reports these results for the selected topic sizes, where the dashed line represents the reference differences measured over the full datasets. As illustrated in Figures~\ref{fig:rq0_2_hotpot} and \ref{fig:rq0_2nq}, using $n=1000$ and $n=2000$ topics for HotpotQA and NQ, respectively, keeps the standard deviation sufficiently low to prevent most zero-crossings, thereby preserving the correct relative ordering of system performance. For a detailed analysis of the remaining topic sizes, we refer the reader to the virtual appendix\footnote{Information at \faicon{github}: \scriptsize \href{https://github.com/IRLab-UDC/Lost-in-the-Evidence-in-RAG-Virtual-Appendix/blob/master/rag\_calibration\_topic\_budget.ipynb}{https://github.com/IRLab-UDC/Lost-in-the-Evidence-in-RAG-Virtual-Appendix/blob/master/rag\_calibration\_topic\_budget.ipynb}}.

\subsection{Controlled and Reproducible Experimental Setup}

Building on the calibration results presented above, we adopt topic sizes that yield low variance and stable ordering trends for each dataset: $n=1000$ topics for HotpotQA and $n=2000$ topics for NQ. We use these fixed topic sets throughout the remainder of the paper to control for topic-sampling noise and enable fair comparisons. 

This controlled evaluation setting supports two subsequent parts of the paper. First, in Section\textcolor{blue}{~\ref{sec:reproducibility}} we revisit and reproduce key prior claims on positional effects and order/size robustness, using the same stable topic sizes to avoid confounding conclusions with sampling variance. Second, in Section\textcolor{blue}{~\ref{sec:controlled-setup}} we conduct our main analyses under the same controlled protocol, isolating how context size and ordering interact with (i) the dataset structure (single-hop vs. multi-hop), (ii) retrieval quality, and (iii) model family and scale.

%\noindent The following subsections present experiments and results corresponding to each research question.

\section{Reproducibility Study of Context Position and Retrieval Robustness}
\label{sec:reproducibility}

In this section, we reproduce two representative lines of work under our controlled evaluation framework. We first test whether long-context positional bias phenomena persist with contemporary LLMs (Subsection\textcolor{blue}{~\ref{subsec:lost-middle}}). We then re-examine recent claims of weak sensitivity to retrieval order and context size in end-to-end RAG, using matched settings and stable topic sizes (Subsection\textcolor{blue}{~\ref{subsec:rag-bloomberg}}). 

\subsection{Reproducing Context Position Effects}
\label{subsec:lost-middle}
% Before deepening into real-world RAG, we aim to study if the well-known effects of \textit{lost in the middle}~\cite{liu-etal-2024-lost-in-the-middle} and \textit{lost but not only in the middle}~\cite{Yu2024InDO} are reproducible using modern, better-performing LLMs. The \textit{lost in the middle} effect describes a U-curve phenomenon where model performance is highest when relevant information is at the beginning or end of the context but drops significantly in the center~\cite{liu-etal-2024-lost-in-the-middle}. In contrast, the \textit{lost but not only in the middle} research suggests that this degradation is more complex, occurring at several different positions across the context rather than being confined strictly to the middle~\cite{Yu2024InDO}.

\begin{figure}[t]
    \centering
    \includegraphics[width=0.95\columnwidth]{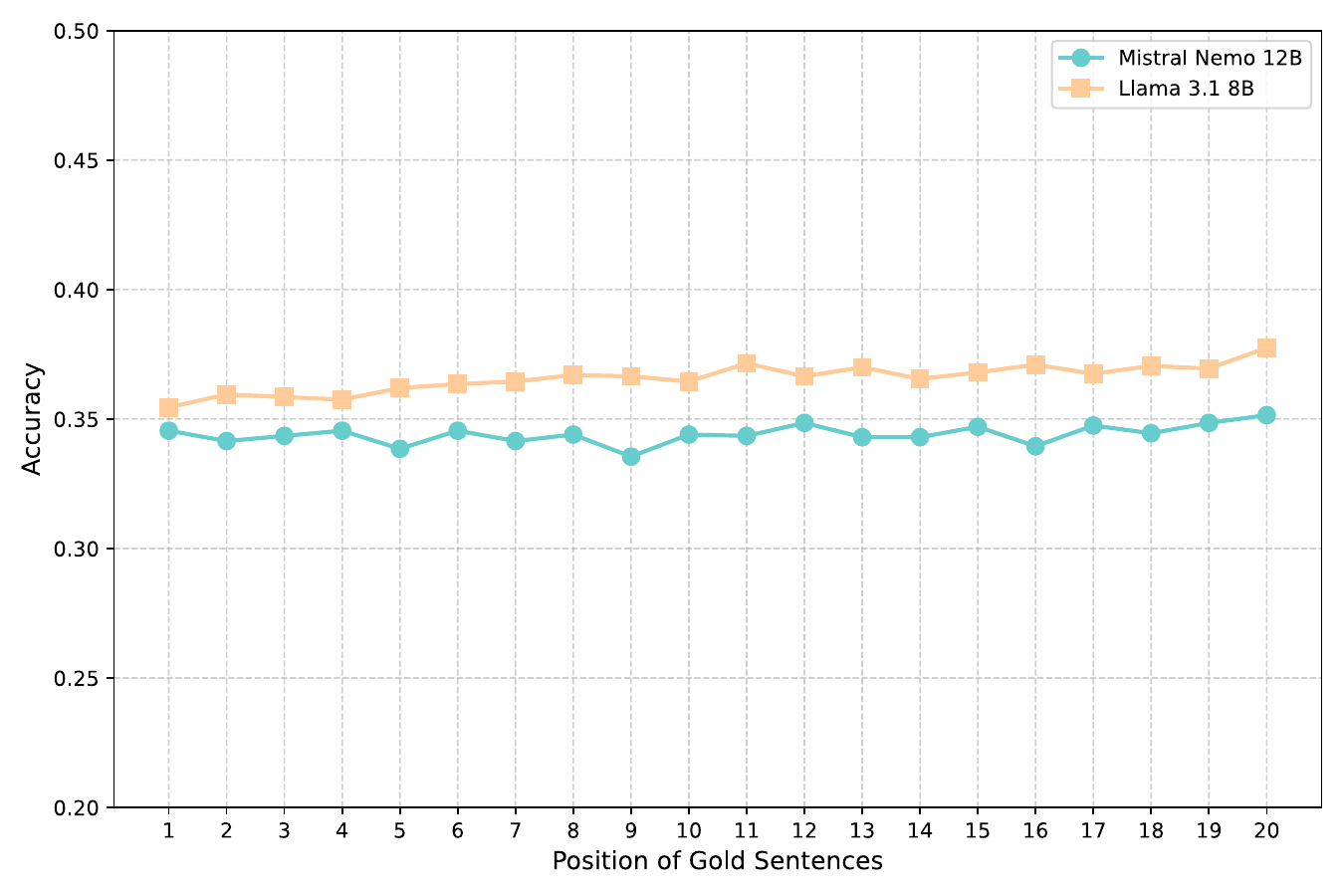}
    \caption{\textbf{\textit{Lost in the middle} reproduction}~\cite{liu-etal-2024-lost-in-the-middle} on AmbigQA~\cite{min2020ambigqa} with LLaMA-3.1:8B and Mistral-NeMo:12B.}
    
    \label{fig:lost-in-the-middle-reproducibility}
\end{figure}

\begin{figure*}[t]
    \centering
    \begin{subfigure}[b]{0.45\textwidth}
        \includegraphics[width=\textwidth]{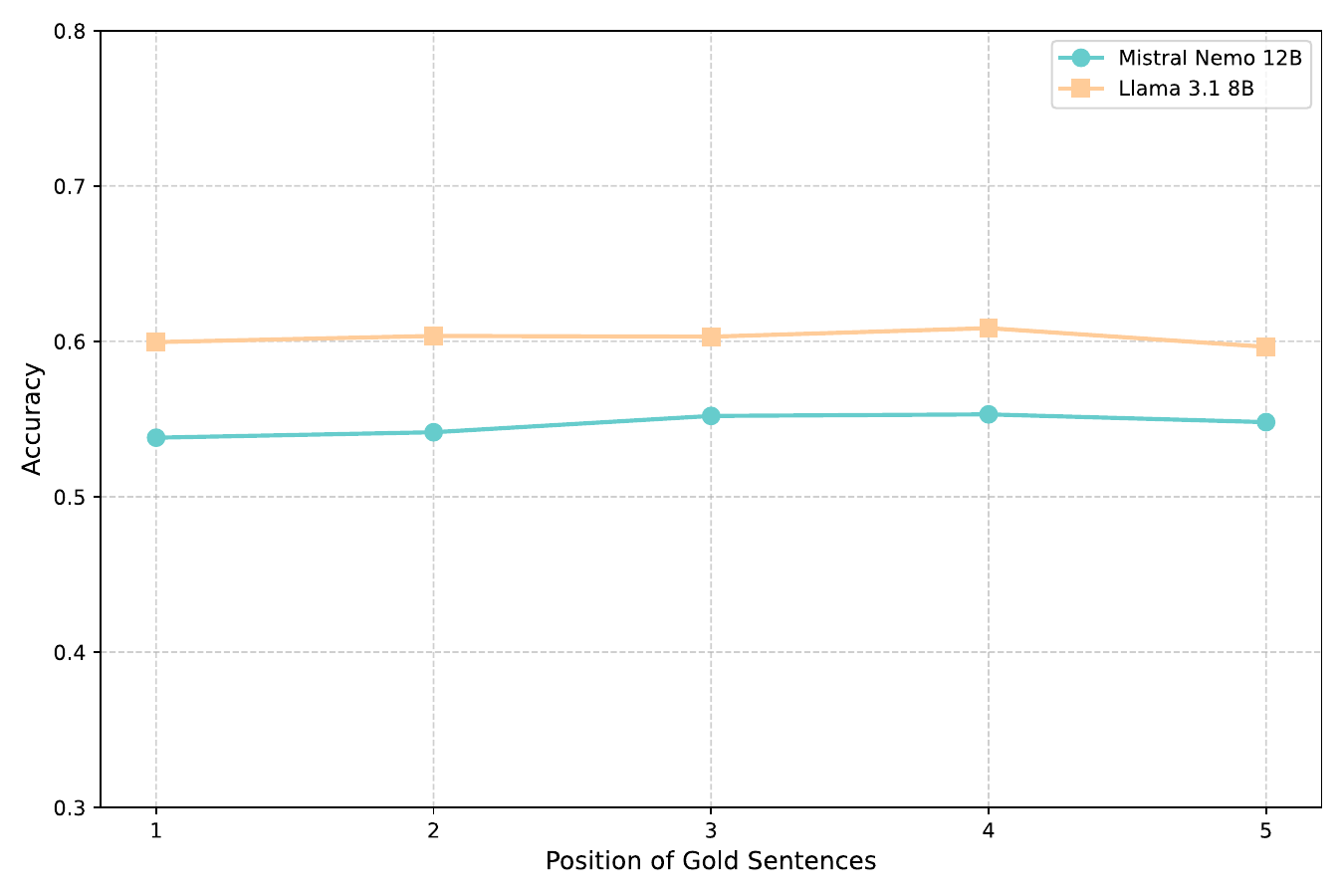}
        \subcaption{NQ Top-5}
    \end{subfigure}
    \begin{subfigure}[b]{0.45\textwidth}
        \includegraphics[width=\textwidth]{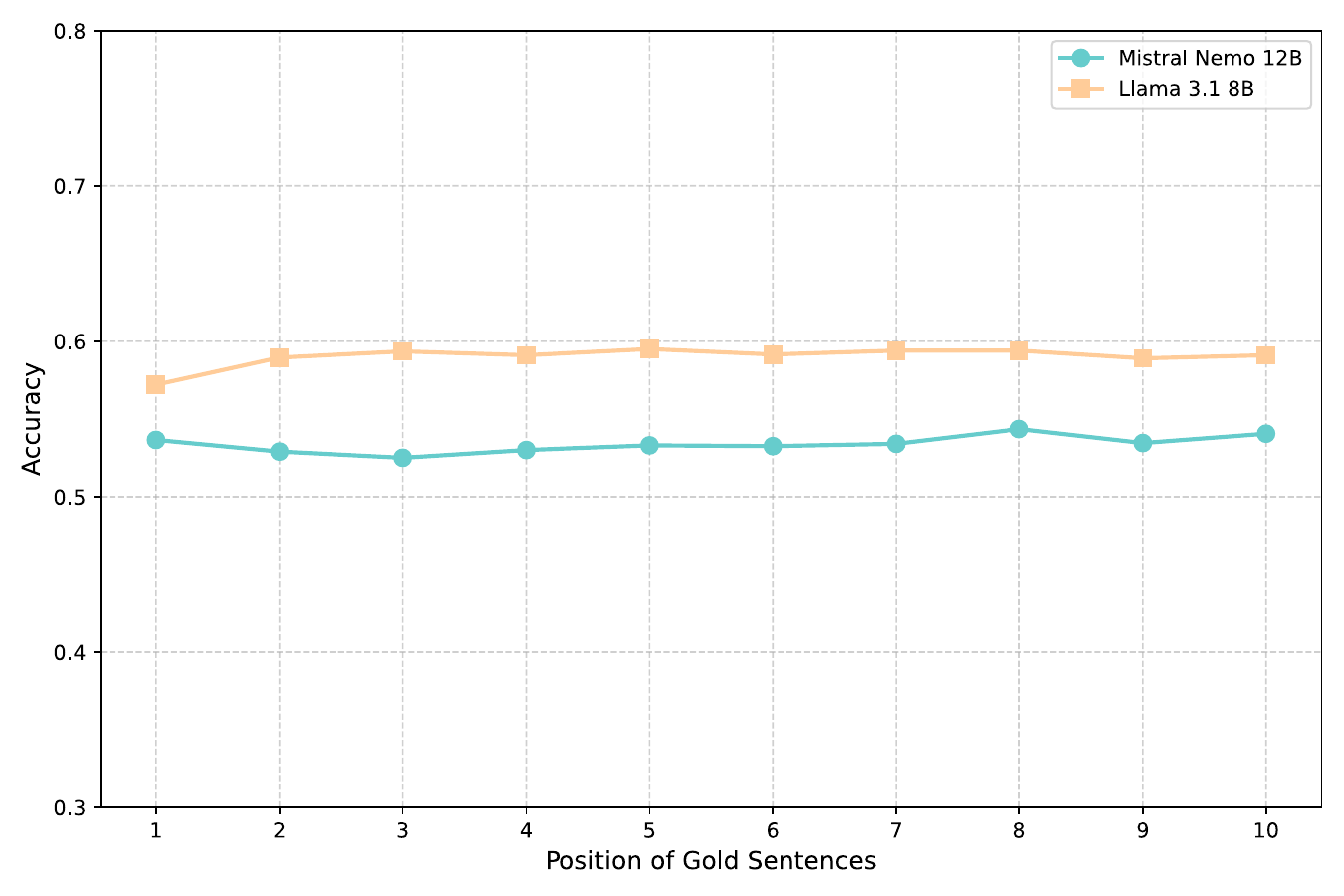}
        \subcaption{NQ Top-10}
    \end{subfigure}
    \begin{subfigure}[b]{0.45\textwidth}
        \includegraphics[width=\textwidth]{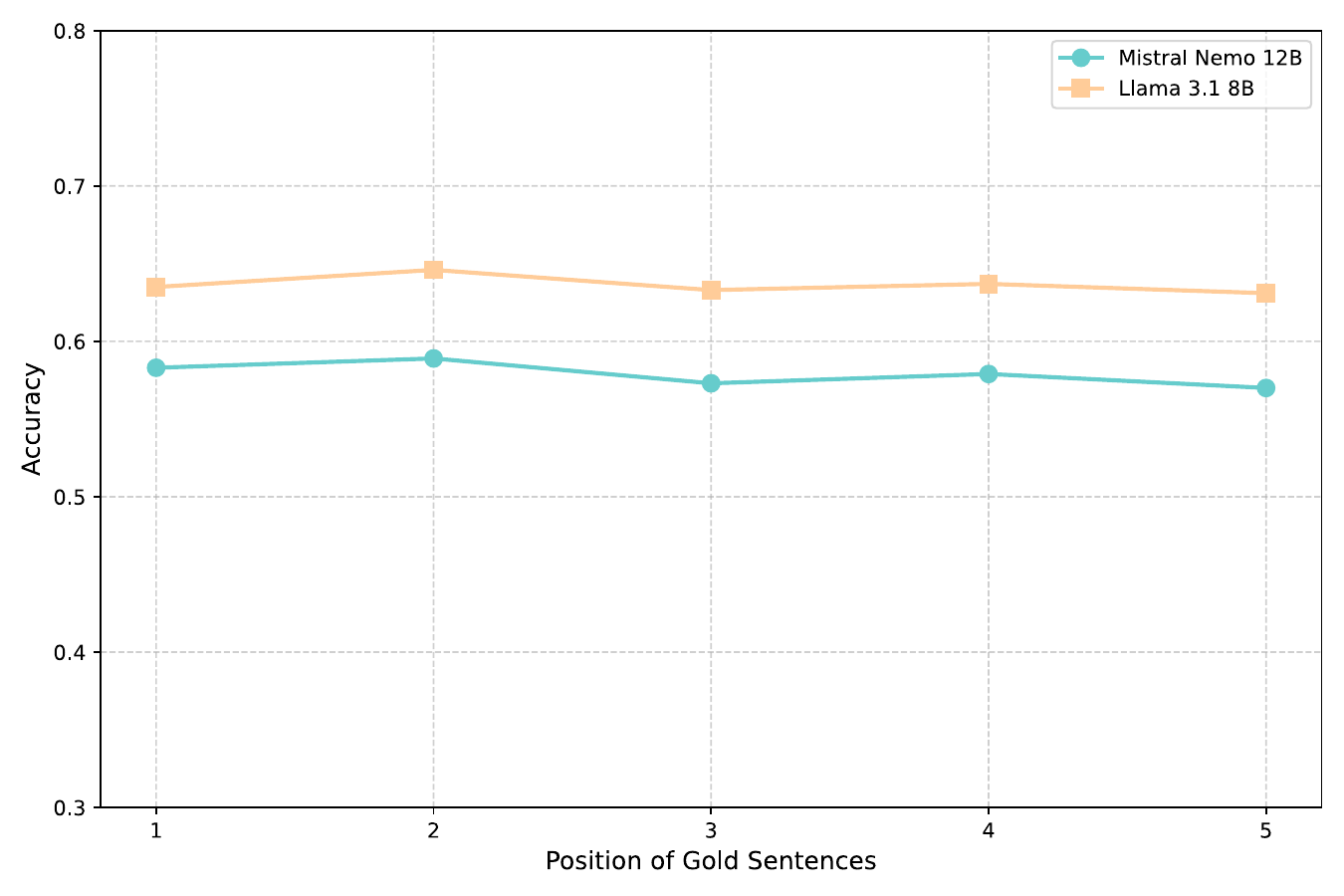}
        \subcaption{HotpotQA Top-8}
    \end{subfigure}
    \begin{subfigure}[b]{0.45\textwidth}
        \includegraphics[width=\textwidth]{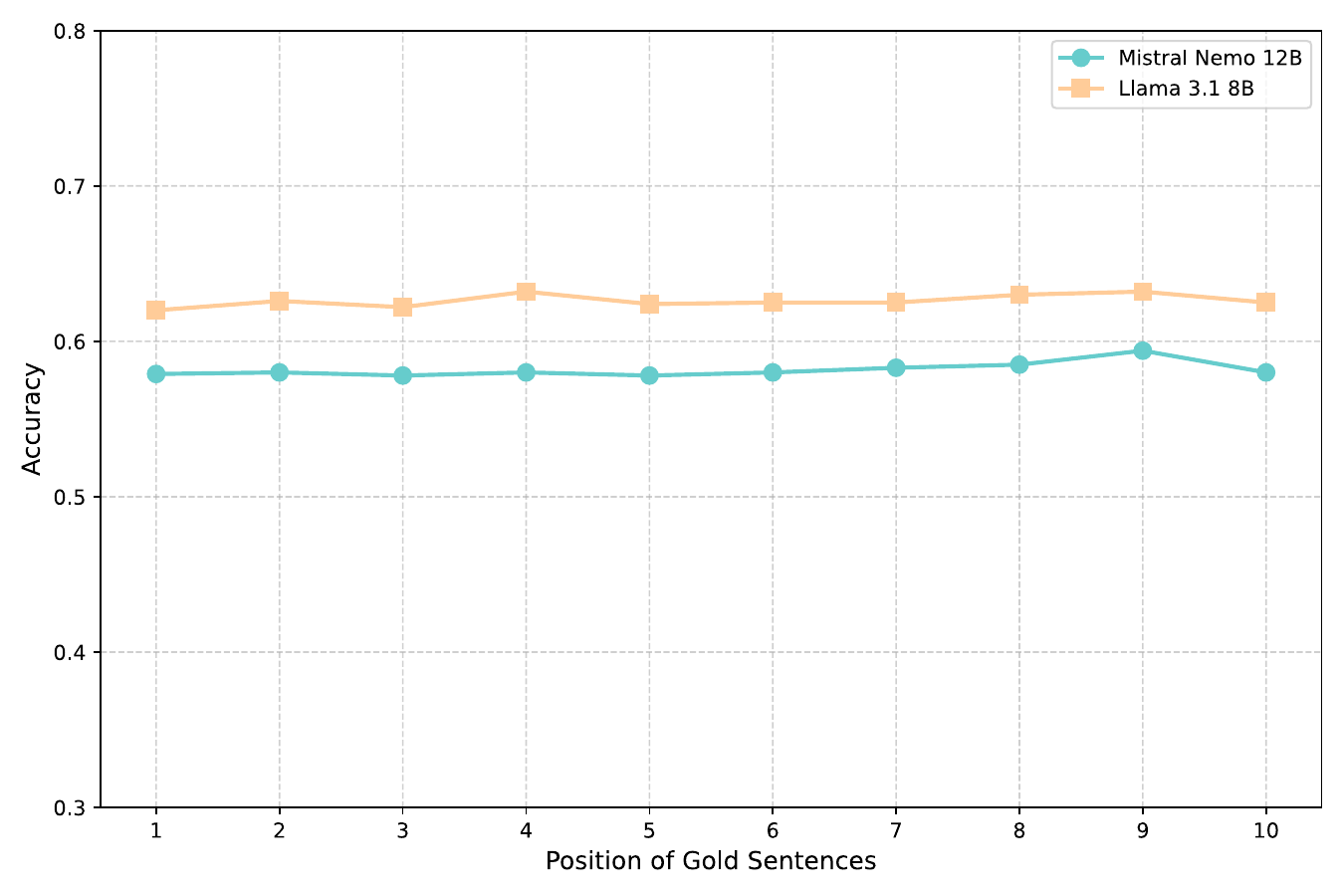}
        \subcaption{HotpotQA Top-13}
    \end{subfigure}
    \caption{\textbf{\textit{Lost but not only in the middle} reproduction}~\cite{Yu2024InDO} on standard NQ and HotpotQA~\cite{kwiatkowski-etal-2019-natural,yang-etal-2018-hotpotqa} using LLaMA-3.1:8B and Mistral-Nemo:12B. For HotpotQA, we evaluate Top-8 and Top-13 instead of standard counts to accommodate the additional context space required for appending all relevant passages, as nearly every query in the dataset contains multiple gold passages.}
    \label{fig:lost-but-not-in-the-middle-reproducibility}
\end{figure*}

We study here whether reported long-context position effects remain visible under our controlled evaluation setting and with contemporary LLMs. The \textit{lost in the middle} phenomenon describes a U-shaped trend: performance is highest when answer-bearing evidence appears near the beginning or end of the context, and drops when the same evidence is placed in the middle~\cite{liu-etal-2024-lost-in-the-middle}. Subsequent work, \textit{lost but not only in the middle}~\cite{Yu2024InDO}, argues that this degradation is not confined to the exact middle and may manifest at multiple positions depending on the prompting and document layout. We revisit both claims using our fixed topic sizes to reduce topic-sampling noise. In these two works, we report accuracy to match the evaluation metric used in both original studies.

\paragraph{\textbf{1) Reproducing \textit{lost in the middle}}~\cite{liu-etal-2024-lost-in-the-middle}} We follow the original study and evaluate on the same dataset, AmbigQA~\cite{min2020ambigqa}. Reproducing the exact protocol is challenging because key dataset handling details are not fully specified in the original paper, and AmbigQA does not provide passage-level annotations that directly support constructing gold-and-distractor contexts in the same way as other standard QA benchmarks. To operationalise the setting, we treat AmbigQA’s reference documents as the gold sources and sample distractor passages from the Natural Questions corpus~\cite{kwiatkowski-etal-2019-natural}. Concretely, we build each context by inserting the gold document into a pool of distractors and sweeping its position across the input. Figure~\ref{fig:lost-in-the-middle-reproducibility} reports results with LLaMA-3.1:8B and Mistral-NeMo:12B, where the x-axis represents the gold documents' position, and the y-axis the accuracy obtained.

From Figure~\ref{fig:lost-in-the-middle-reproducibility}, we observe two main differences from the originally reported behaviour. First, we do not recover a clear U-shaped curve: accuracy remains comparatively flat across positions, with only a slight upward trend. Second, absolute scores differ from those reported in~\cite{liu-etal-2024-lost-in-the-middle}, which is consistent with small but consequential differences in dataset processing and document construction (the original paper notes using additional support from the dataset authors). Together, these results suggest that the originally observed position effect is not straightforwardly reproducible, considering our setting and a modern LLM.

\paragraph{\textbf{2) Reproducing \textit{lost but not only in the middle}}~\cite{Yu2024InDO}} We next reproduce the study of Yu et al.~\cite{Yu2024InDO}. The original experiments were conducted on KILT (formatted versions of NQ and HotpotQA). In our setting, we instead use their standard releases~\cite{kwiatkowski-etal-2019-natural,yang-etal-2018-hotpotqa}, which are the default format in most QA and RAG pipelines. Moreover, both AmbigQA (in the previous experiment) and HotpotQA can contain multiple relevant documents/passages. Since prior work does not specify how multi-evidence cases are positioned, we adopted here a simple convention: the x-axis position $i$ indicates where the first relevant document/passage is placed, and any additional relevant documents are inserted immediately after it. This ensures keeping the evidence grouped and avoids mixing evidence with distractors in a way that would confound the positional sweep. We run the positional sweep protocol using LLaMA-3.1:8B and Mistral-Nemo:12B. Figure~\ref{fig:lost-but-not-in-the-middle-reproducibility} shows that performance is again nearly flat across placements, providing little evidence of degradation at specific positions for modern models under our setup.

% \begin{figure*}[t]
%     \centering
%     \begin{subfigure}[b]{0.48\textwidth}
%         \includegraphics[width=\textwidth]{new_figures/nq_5_total_docs_comparison.pdf}
%     \end{subfigure}
%     \begin{subfigure}[b]{0.48\textwidth}
%         \includegraphics[width=\textwidth]{new_figures/nq_10_total_docs_comparison.pdf}
%     \end{subfigure}
%     \begin{subfigure}[b]{0.48\textwidth}
%         \includegraphics[width=\textwidth]{new_figures/hotpotqa_7_total_docs_comparison.pdf}
%     \end{subfigure}
%     \begin{subfigure}[b]{0.48\textwidth}
%         \includegraphics[width=\textwidth]{new_figures/hotpotqa_12_total_docs_comparison.pdf}
%     \end{subfigure}
%     \caption{Caption}
%     \label{fig:placeholder}
% \end{figure*}

\subsection{Reproducing Order and Size Robustness Under Retrieval}
\label{subsec:rag-bloomberg}

\begin{figure}[t]
    \centering
    \begin{subfigure}[b]{0.49\textwidth}
        \centering
        \includegraphics[width=\textwidth]{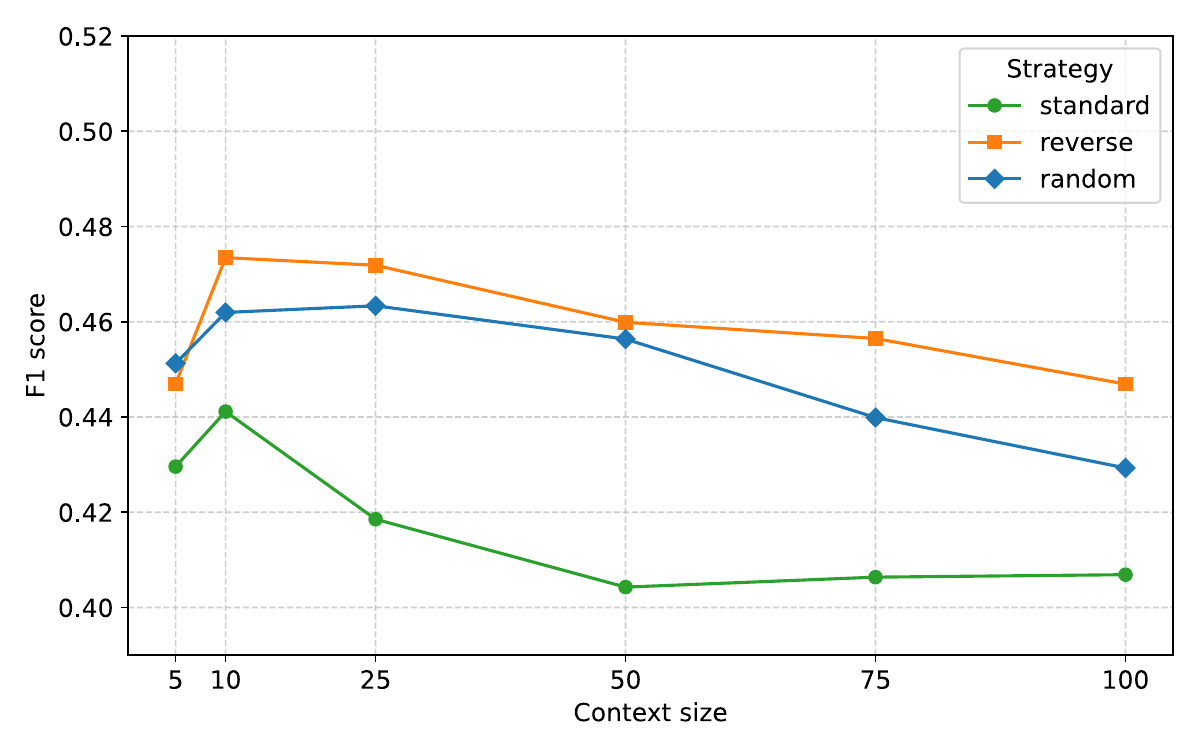}
        \caption{HotpotQA (500 topics)}
        \label{fig:rq2_hotpot}
    \end{subfigure}
    \hfill
    \begin{subfigure}[b]{0.49\textwidth}
        \centering
        \includegraphics[width=\textwidth]{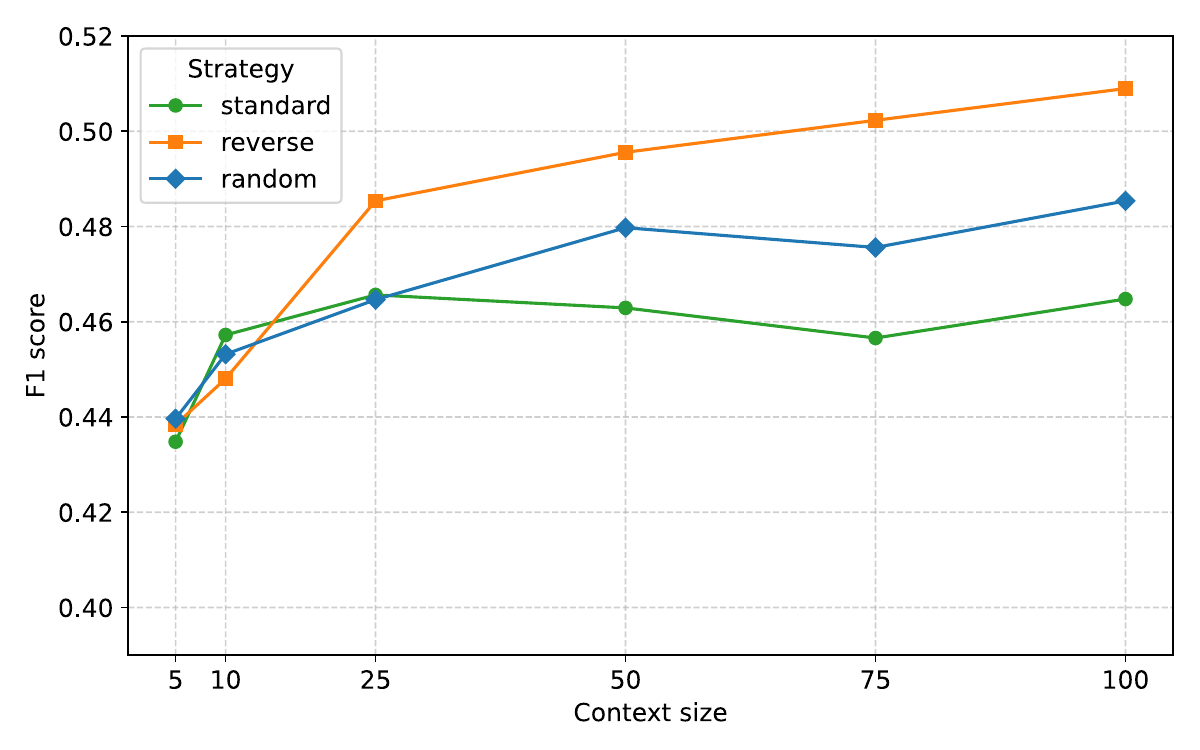}
        \caption{NQ (500 topics)}
        \label{fig:rq2_nq}
    \end{subfigure}
    \caption{Context ordering and size effects using LLaMA-3.1:8B and 500 random topics. Results differ from those in \cite{cao-bloomberg-2025}, suggesting that context ordering influences RAG performance more than previously observed.}
    \label{fig:rq2}
\end{figure}

Both prior studies reproduced above deliberately guarantee that the gold evidence is present in the context and vary only its location with a synthetic distractor set. While useful for isolating positional sensitivity, they are optimistic relative to real-world RAG scenarios, where the retriever may fail to return all gold documents, and relevance is graded rather than binary. For this reason, we now shift to a more realistic setting, in which context order and size emerge from imperfect retrieval. Our goal here is to reproduce recent claims that RAG performance is \emph{weakly sensitive} to both (i) the number of retrieved documents and (ii) the order in which those documents are concatenated into the prompt.

\paragraph{\textbf{Target study}} In a recent industry paper, Cao et al.~\cite{cao-bloomberg-2025} evaluate \emph{retrieval robustness} in practical RAG settings using a benchmark of 1{,}500 open-domain questions (500 each from NQ, HotpotQA, and ASQA) with Wikipedia retrieval. They vary the retrieval depth $k\in\{5,10,25,50,75,100\}$ and compare three ordering schemes applied to the same top-$k$ set: \emph{original rank}, \emph{reversed rank}, and \emph{random shuffle}. Their main conclusion is that average performance is often relatively stable across orderings and retrieval depths, even though they also report residual \emph{sample-level} trade-offs. Importantly, they evaluate correctness using an LLM-as-a-judge (LLaMA-3.3:70B) rather than string-match style metrics (F1-token).

\paragraph{\textbf{Our reproduction setup}} To mirror this protocol as closely as possible while keeping our analysis aligned with the rest of the paper, we focus on the two datasets that we use throughout: NQ and HotpotQA. We evaluate LLaMA-3.1:8B under the same $k$ grid and the same three ordering schemes (\emph{original}, \emph{reversed}, \emph{random}) on 500 randomly sampled topics per dataset, matching the per-dataset topic budget used by Cao et al.~\cite{cao-bloomberg-2025}. As in the target study, each condition uses the same retrieved top-$k$ set; only the concatenation order changes.

While Cao et al.~\cite{cao-bloomberg-2025} rely on an LLM-based judge, we report token-level F1 as our primary metric. To verify that this choice does not change the qualitative ordering conclusions, we ran a pilot experiment on a subset of $n=\textit{250}$ topics from NQ and HotpotQA, evaluating the same conditions with both F1 and an LLM judge. Across all ($top$-$k$,\text{order}) settings, the two scorers exhibited near-identical ordering gaps, indicating that the relative separation between ordering schemes is essentially unchanged. Given this agreement, and since F1 is the official benchmark metric and more reproducible, we use F1 throughout for reproducibility and comparability across sections.

Figure~\ref{fig:rq2} shows the reproduced curves for NQ and HotpotQA (500 topics each), plotting the three ordering strategies, with the x-axis showing the context size (number of retrieved passages, $k\!\in\!\{5,10,25,50,75,100\}$) and the y-axis the F1-score. In contrast to the strong average stability reported in \cite{cao-bloomberg-2025}, we observe measurable sensitivity to both context size and ordering, with differences that are small at low $k$ but become more apparent as $k$ grows. Even under a matched configuration (same datasets, model family, 500-topic samples, and the three orderings), reproducing the original average stability trend proves difficult. 

% This pattern is consistent with the intuition that longer contexts amplify positional biases and increase the burden of evidence selection within the prompt.

Looking at these results, we can see that the more fundamental issue is that both settings operate with relatively small topic samples. When per-query variance is high, and the marginal gains between orders are modest, conclusions about ``order robustness'' can depend strongly on which topics are sampled. This motivates our controlled evaluation framework (Section \ref{sec:towards-stable-evaluation}). In the next section, we adopt the calibrated topic sizes identified earlier and re-evaluate ordering and context-size effects under stable topic sets, so that subsequent comparisons across models, retrievers, and rerankers are not confounded by topic-sampling noise.

\section{Controlled RAG Analysis: Ordering, Depth, Retrieval Quality, and Models}
\label{sec:controlled-setup}

Having established a stable evaluation protocol and revisited key prior claims, we now conduct targeted analyses to characterise when and why context size and ordering matter in practical RAG pipelines. Unlike idealised position-sweep settings, the evidence available to the generator is mediated by imperfect retrieval, and the marginal utility of adding more context depends on both the ranking quality and how the model allocates attention across long inputs. Using the fixed topic sizes for NQ and HotpotQA identified in Section~\ref{sec:towards-stable-evaluation}, we study four research questions:

\begin{itemize}
    \item \textbf{RQ1.} How do different document ordering strategies and context sizes affect QA performance? (Subsection \textcolor{blue}{~\ref{subsec:rq1}})
    
    \item \textbf{RQ2.} How much performance is attributable to the LLM versus the retrieved evidence quality, as estimated by closed-book and oracle contexts?  (Subsection \textcolor{blue}{~\ref{subsec:rq2-3}})
    
    \item \textbf{RQ3.} How does retrieval quality (BM25 alone versus BM25 with dense reranking) interact with context ordering and size? (Subsection \textcolor{blue}{~\ref{subsec:rq2-3}})
    \item \textbf{RQ4.} How do model family and scale influence sensitivity to ordering and context size? (Subsection \textcolor{blue}{~\ref{subsec:rq4}})
\end{itemize}

\subsection{Exploring Context Size and Ordering Effects (RQ1)}
\label{subsec:rq1}

% This is for \item \textbf{RQ4.} How do different document ordering strategies and context sizes affect model performance?

\begin{figure}[t]
    \centering
    \begin{subfigure}[b]{0.49\textwidth}
        \centering
        \includegraphics[width=\textwidth]{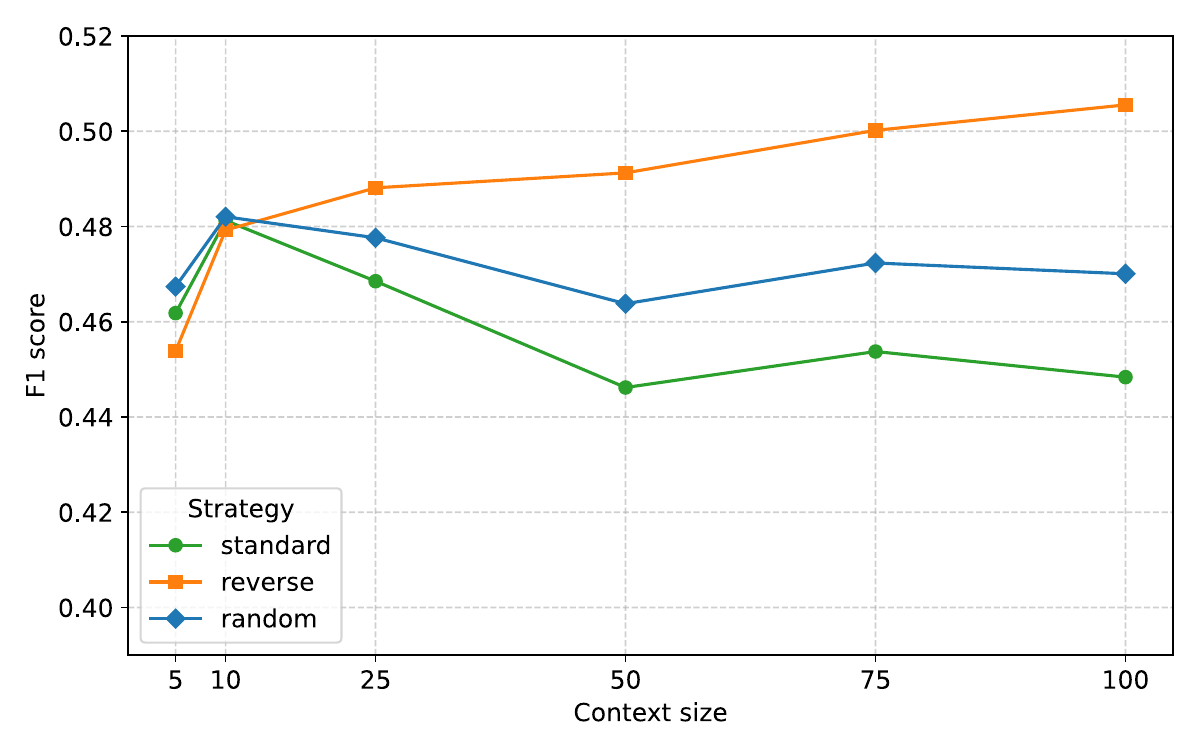}
        \caption{HotpotQA (1000 topics)}
        \label{fig:rq4_hotpot}
    \end{subfigure}
    \hfill
    \begin{subfigure}[b]{0.49\textwidth}
        \centering
        \includegraphics[width=\textwidth]{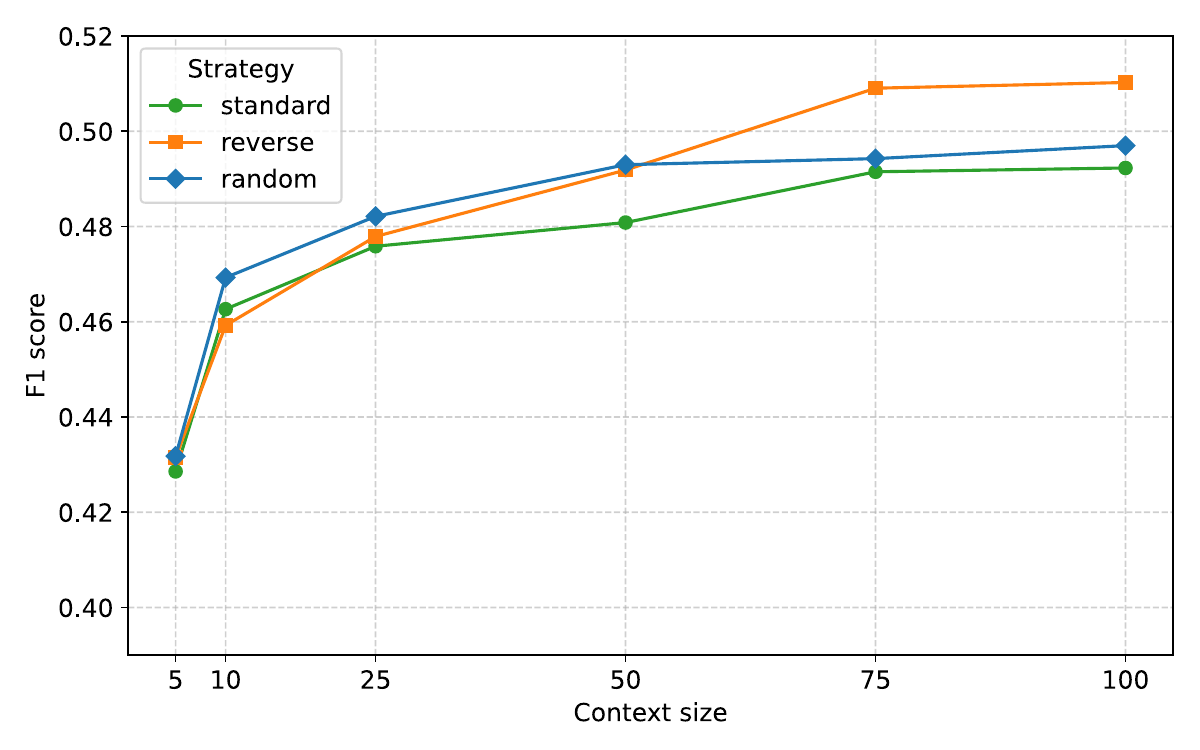}
        \caption{NQ (2000 topics)}
        \label{fig:rq4_nq}
    \end{subfigure}
    \caption{Effect of document ordering and context size on model performance for HotpotQA and NQ. Results show the impact of different ordering strategies and context size.}
    \label{fig:rq4}
\end{figure}

This experiment investigates how context size and document ordering influence RAG performance under the controlled setup defined earlier ($n=1000$ topics for HotpotQA and $n=2000$ for NQ). We vary two parameters: $i$) the number of retrieved passages and $ii$) their ordering within the prompt to examine their combined effect on information utilisation and answer accuracy. We again compare the three ordering schemes: \textit{standard}, \textit{reverse}, and \textit{random}. Each strategy is evaluated across context sizes (ranging from $k\!\in\!\{5,10,25,50,75,100\}$), allowing us to assess the trade-off between context length and retrieval order. 

Figure \ref{fig:rq4} shows the results for this experiment, plotting F1 on the y-axis against the context size on the x-axis. The results display a dataset-dependent behaviour. On HotpotQA, model performance is more sensitive to ordering: the reverse scheme increasingly outperforms other strategies at larger contexts, indicating that placing stronger passages later in the prompt helps counter position sensitivity in multi-hop settings. In contrast, on NQ, performance rises with $k$ and remains comparatively order-stable, suggesting that single-hop questions benefit primarily from additional recall rather than from a particular order.

Examining the overall trends and addressing RQ1, we demonstrate that the effectiveness of additional evidence depends on its positioning in the prompt in multi-hop settings. Additional passages help only when high-value passages are placed in positions the model prefers. These observations are consistent with prior work~\cite{izacard-grave-2021-leveraging,Yu2024InDO}. On single-hop QA, multi-passage generative readers such as Fusion-in-Decoder (FiD) show that supplying more relevant passages generally improves accuracy~\cite{izacard-grave-2021-leveraging}. Conversely, when contexts become long, order-aware layouts such as OP-RAG~\cite{Yu2024InDO}, which preserve intra-document chunk order instead of relevance, improve answer quality and exhibit an inverted-U relationship between performance and context size~\cite{Yu2024InDO}.

\subsection{Impact of Retrieval Quality and Reranking (RQ2 and RQ3)}
\label{subsec:rq2-3}
% This of for the \item \textbf{RQ5.} To what extent can performance be attributed to the underlying LLM versus the quality of the retrieved context, as illustrated by an oracle ranking and the closed-book performance?

\paragraph{\textbf{RQ2.}} To quantify how much performance derives from the retrieved context versus the LLM itself, we propose and evaluate a wide range of configurations that progressively increase evidence quality while holding the LLM fixed (LLaMA-3.1:8B). In this analysis, we focus on HotpotQA because it provides \emph{gold passage} and \emph{gold sentence} annotations for supporting evidence, which are required to construct oracle contexts and to separate passage-level from sentence-level upper bounds. We evaluate the next configurations:

\begin{itemize}
    \item \textbf{Closed-book (\emph{no retrieval})}: the model answers without any external passages, our lower bound on knowledge and reasoning without context.
    
    \item \textbf{BM25 retrieval}: standard top-$k$ BM25, evaluated under the three prompt orders to capture order sensitivity: standard, reverse, and random.
    
    \item \textbf{Oracle-passages:} an upper-bound selection that includes all gold-relevant passages as context.
    
    \item \textbf{Oracle-sents:} a stricter upper bound that includes only the exact sentences from the gold-relevant passages that contain the facts needed to answer.

    \item \textbf{Oracle-passages+BM25 (standard)}: a strategy that places first all gold-relevant passages, then fills remaining context slots by using the top non-relevant BM25-ranked passages in the standard ranking order up to size $k$.
    
    \item \textbf{Oracle-passages+BM25 (reverse)}:  places the BM25-selected non-relevant passages first, in ascending BM25 score order, and appends the gold-relevant passages at the end.

\end{itemize}

\begin{figure}[t]
    \centering
    \includegraphics[width=\columnwidth]{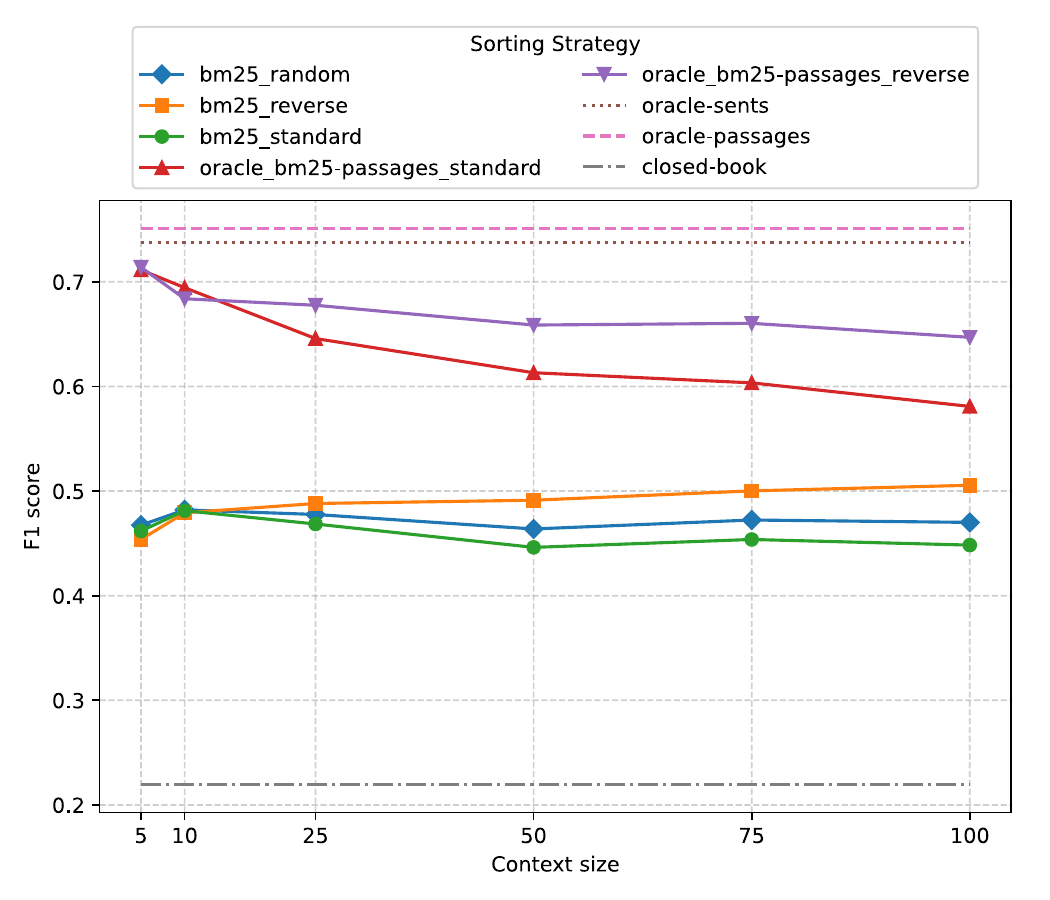}
    \caption{Comparison of closed-book, standard retrieval (BM25), and oracle ranking results on HotpotQA. The oracle ranking illustrates the upper bound achievable with a perfect ordering, showing the relative contributions of the LLM and the retrieval component to overall RAG performance.}
    \label{fig:rq5_hotpotqa}
\end{figure}

% \textcolor{red}{anxo: quizas me he explayado mucho comentado los resultados aqui de las 8 estrategias, se puede reducir. Ademas, pequeña justificacion de por que solo usamos hotpotQA aqui -> jorge: si, hay que explicar que solo usamos ese dataset porque es el que tiene labelleados los passages/sentencias/evidencias que responden a las preguntas (por lo que yo vi natural questions no tiene eso) -> en el caso de nq por lo que veo se podría llegar a usar la long answer (según lo que vi es un párrafo de wikipedia que responde)}

Figure~\ref{fig:rq5_hotpotqa} reports F1 on the y-axis versus context size ($k$) on the x-axis for HotpotQA across the eight configurations. Looking at the results, two main trends emerge. First, the closed-book baseline is the weakest, underscoring that model base knowledge is insufficient for this multi-hop setting. Second, both oracle-passages (all gold-relevant passages) and oracle-sents (only the sentences containing the answer) form the upper bound, clustering a bit above 0.70 F1 across $k$. This isolates the upper bound attributable to evidence quality, not the LLM capacity. Within BM25 retrieval, the three ordering schemes start close together at small $k$, then diverge mildly as $k$ grows: reverse pulls ahead of standard and random for larger contexts, and all three curves stabilise after 50 passages. This aligns with prior position effects: positioning the key passages at the end is beneficial in long contexts. Moreover, the gap between BM25 and oracle results quantifies the remaining potential for retrieval and ordering optimisation.

In Figure~\ref{fig:rq5_hotpotqa}, we can also see that the oracle with BM25-noise introduction conditions further differentiates ordering from evidence quality. For Oracle+BM25 (standard), performance decreases as $k$ increases: once the oracle content is present, adding more BM25 material in standard rank adds noise and displaces the crucial evidence from border positions. In contrast, Oracle+BM25 (reverse) consistently dominates its standard counterpart and remains the better approach for all $k \ge 10$, suggesting that even when gold passages are present, placing them at the tail better aligns with the model’s reasoning.

% Figure~\ref{fig:rq5_hotpotqa} presents results on HotpotQA. The closed-book baseline shows the lowest performance, highlighting the model’s limited capacity to retrieve factual information from parameters alone. BM25 retrieval substantially improves F1 scores, while the oracle ranking demonstrates the maximum gain under perfect ordering. This also illustrates the ``real'' importance of document ordering: improvements from the reverse strategy are substantial relative to the standard ordering, showing that optimal ranking can significantly influence performance. The gap between BM25 and oracle results quantifies the remaining potential for retrieval and ordering optimisation.

\begin{figure}[t]
    \centering
    \begin{subfigure}[b]{0.49\textwidth}
        \centering
        \includegraphics[width=\textwidth]{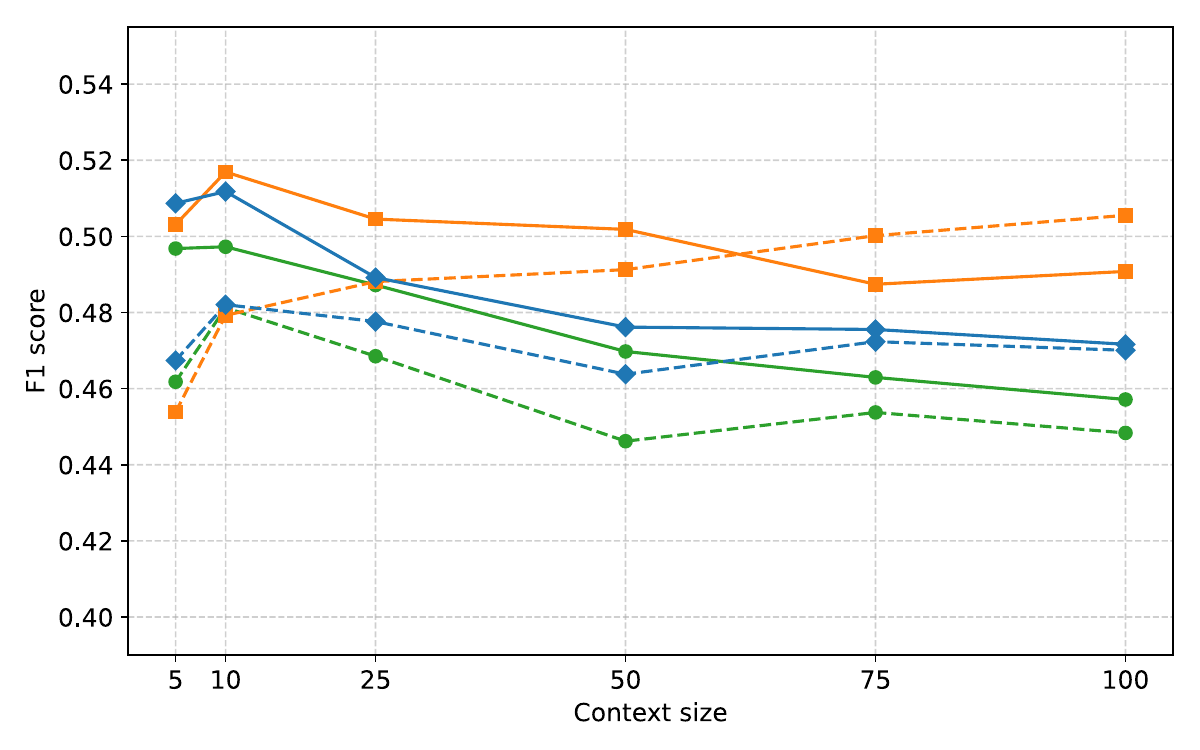}
        \caption{HotpotQA (1000 topics)}
        \label{fig:rq6_hotpot}
    \end{subfigure}
    \hfill
    \begin{subfigure}[b]{0.49\textwidth}
        \centering
        \includegraphics[width=\textwidth]{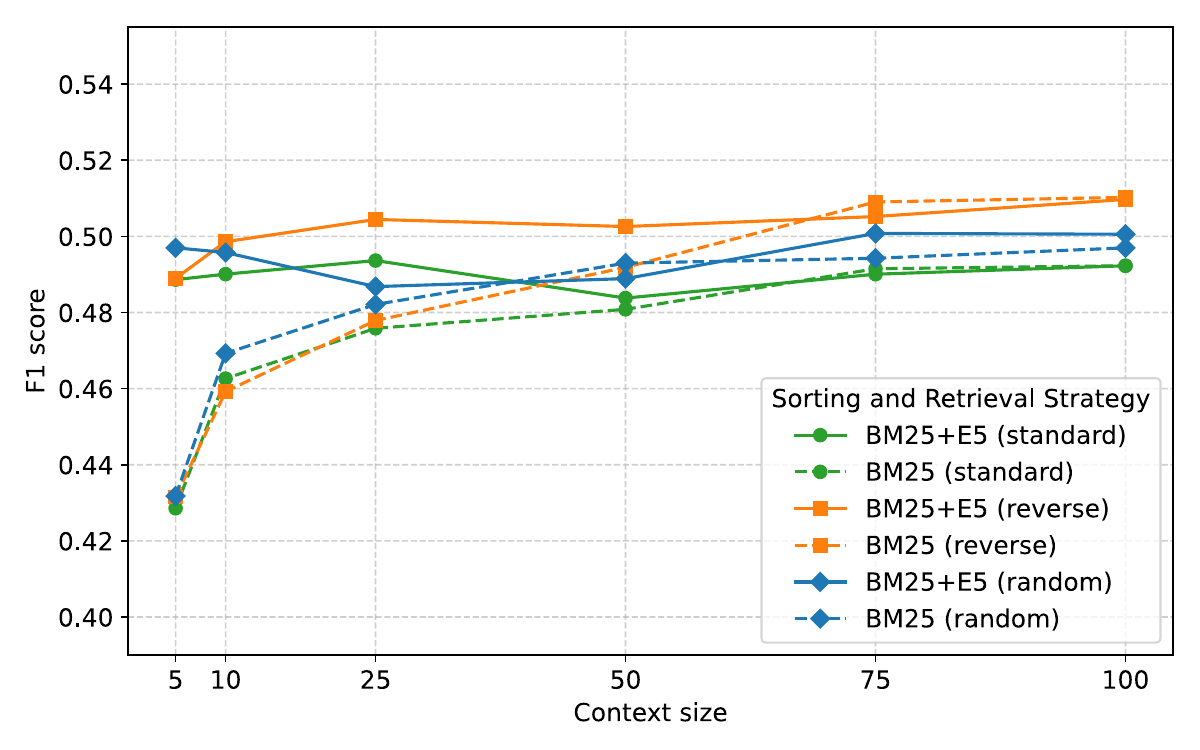}
        \caption{NQ (2000 topics)}
        \label{fig:rq6_nq}
    \end{subfigure}
    \caption{Impact of retrieval model on RAG performance. Results illustrate the performance of BM25 and dense reranking, showing how improved retrieval interacts with ordering strategies, influencing the consistency of generated answers.}
    \label{fig:rq6}
\end{figure}

% This answers \item \textbf{RQ6.} How does retrieval quality (BM25 alone versus BM25 with dense reranking) interact with context ordering?
\paragraph{\textbf{RQ3.}} We now study how retrieval quality interacts with ordering and context size. Beyond BM25, we add a reranking stage that first retrieves a candidate pool with BM25 and then re-scores the top 100 of those candidates using E5 embeddings~\cite{wang2024multilingual} with cosine similarity, finally presenting the top $k$ to the LLM. We choose E5 because it is a strong, open sentence-embedding model trained for retrieval-style objectives, offers solid zero-shot performance across domains, is efficient for large candidate pools, and improves reproducibility ~\cite{wang2024multilingual}. 

Figure~\ref{fig:rq6} plots F1 (y-axis) against context size $k$ (x-axis) for BM25 and reranking (BM25+E5), each under the three ordering schemes (standard, reverse, random). In these new results, with reranking, small contexts ($k\in\{5,10\}$) achieve noticeably higher performance than BM25 alone on both datasets, indicating that higher-quality top-$k$ reduces dependency on long prompts. As $k$ grows, HotpotQA shows a decline under reranking, as lower-quality passages added to the end of the prompt dilute the multi-hop structure. However, NQ performance as $k$ grows remains roughly flat. For long contexts ($k \ge 50$ ), curves under BM25+E5 tend to approach the BM25 levels, and the gaps between the three order strategies  narrow. This suggests that better evidence quality decreases order sensitivity and that further gains are limited by prompt length rather than ranking.

Taken together, these trends indicate that stronger retrieval favours shorter contexts: when the first few passages are high quality, adding more tends to offer diminishing or negative returns (especially on multi-hop), and the specific prompt order matters less. Conversely, under weaker retrieval, longer contexts and their order remain consequential. This quantifies the interplay in RQ4: improvements from reranking reduce the need for long contexts and reduce ordering effects, whereas BM25 alone benefits more from reverse ordering at higher $k$.

% Next, we evaluate how retrieval model choice interacts with ordering and context size. Figure~\ref{fig:rq6} shows the performance of LLaMA-3.1 8B using a reranking setup during retrieval ($BM25$ >> $E5$). Unlike BM25 alone, smaller contexts (5–10 documents) yield higher performance, while adding more documents tends to decrease the performance (HotpotQA) or remain stable (NQ). This suggests that higher-quality retrieval reduces the need for longer contexts and lessens the impact of document ordering.

% Together, these results demonstrate that retrieval quality is the main determinant of RAG effectiveness. While ordering matters under weaker retrieval methods, its influence diminishes as retrieval and reranking improve, shifting the performance bottleneck from document selection to model comprehension. This directly addresses RQ5 and RQ6 by quantifying the interplay between retrieval quality, context length, and ordering.

\subsection{Trends Across Model Sizes and Architectures (RQ4)}
\label{subsec:rq4}

\begin{figure}[t]
    \centering
    \begin{subfigure}[b]{0.49\textwidth}
        \centering
        \includegraphics[width=\textwidth]{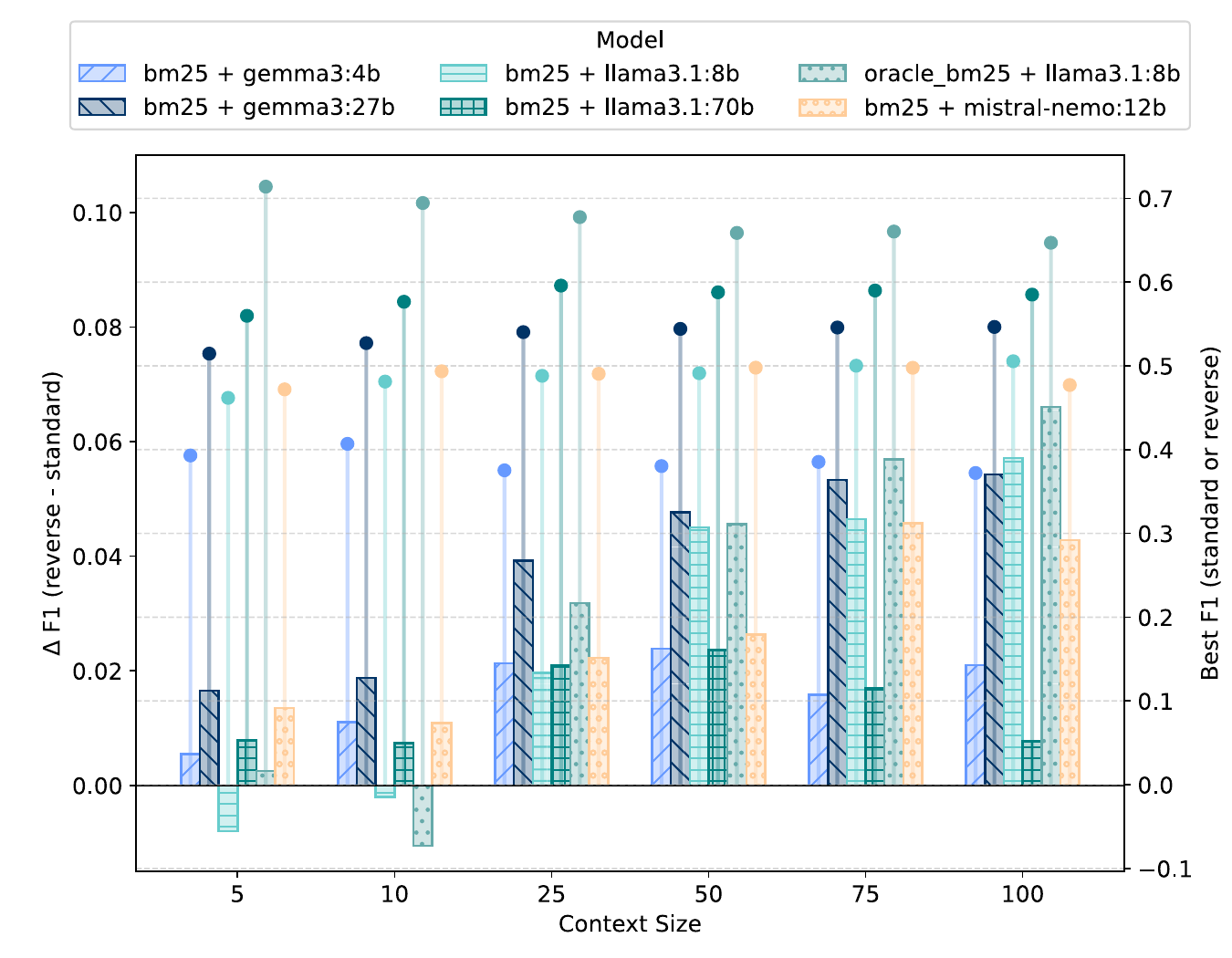}
        \caption{HotpotQA (1000 topics)}
        \label{fig:rq7_hotpot}
    \end{subfigure}
    \hfill
    \begin{subfigure}[b]{0.49\textwidth}
        \centering
        \includegraphics[width=\textwidth]{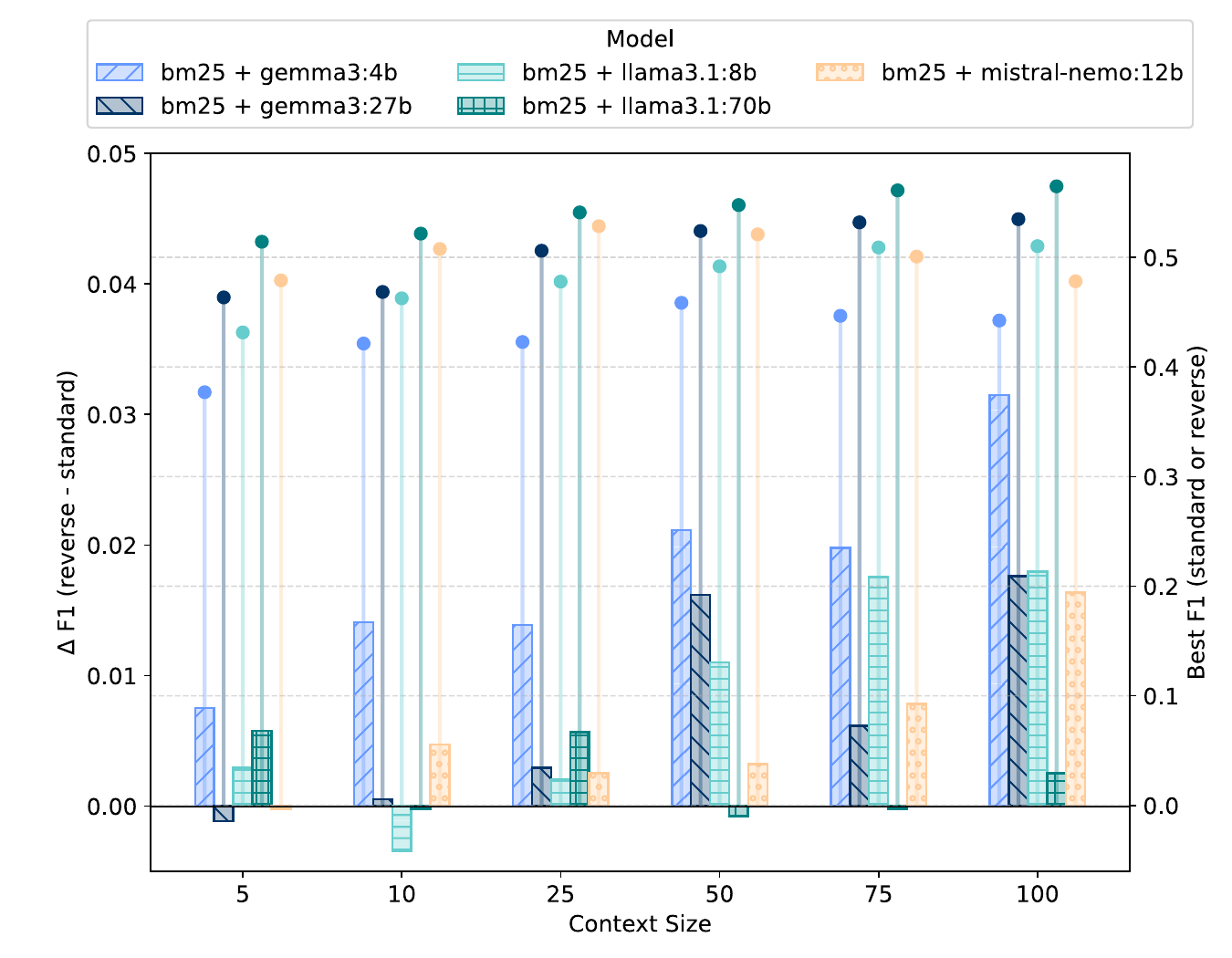}
        \caption{NQ (2000 topics)}
        \label{fig:rq7_nq}
    \end{subfigure}
    \caption{Trends in model robustness to context ordering across different architectures and sizes. The left y-axis and bars show the variations in F1 (\(\Delta\)F1) between reverse and standard ordering, while the right y-axis and lollipops indicate the best F1 achieved by each model, regardless of ordering.}
    \label{fig:rq7}
\end{figure}

All prior experiments use LLaMA-3.1~8B, which stands as a strong, widely adopted baseline that offers competitive QA performance at manageable cost. We now test whether ordering and context-size effects generalise across different LLM families and scales. To summarise our study without overcrowding plots, for all the models, we report \(\Delta\)F1 \(=\) F1\(_{\text{reverse}}\) \(-\) F1\(_{\text{standard}}\) for the different context sizes. This highlights the marginal effect of reversing the order at each context size and facilitates cross-model comparison. Figure~\ref{fig:rq7} shows the results difference of reverse and standard ordering strategies for HotpotQA (Figure~\ref{fig:rq7_hotpot}) and NQ (Figure~\ref{fig:rq7_nq}) across all architectures and sizes. Looking at the overall results, different key patterns emerge.

% All previous experiments used a single LLaMA model (Llama 3.1 8B). Here, we examine whether the observed effects of context size and document ordering generalise across different model families and sizes. 

% To summarise these results without overcrowding the figures, we compute the delta in F1 scores between the \textit{reverse} and \textit{standard} ordering strategies. This approach highlights the relative impact of document ordering and context size. Moreover, it allows us to compare trends in sensitivity across models efficiently, revealing how variability depends on architecture and model capacity.

First, robustness to document ordering is strongly dataset dependent. In HotpotQA, Figure~\ref{fig:rq7_hotpot} shows a consistent increase in \(\Delta\)F1 with \(k\) for most models: reversing the order helps more as contexts become longer, in line with the position effects observed earlier. Larger models (e.g., LLaMA-3.1:70B) tend to have \(\Delta\)F1 values that are closer to zero and smoother, indicating reduced sensitivity. Mid-sized models (e.g., LLaMA-3.1:8B, Mistral-Nemo:12B) exhibit larger positive oscillations at higher \(k\). Interestingly, the smallest model (Gemma-3:4B) remains relatively stable around small positive values. For NQ (Figure~\ref{fig:rq7_nq}), \(\Delta\)F1 is generally small near zero across \(k\), reflecting the order stability seen in the main results. Smaller models (e.g., Gemma-3:4B) can now show more variability, while larger models produce lower values. This suggests that, for predominantly single-hop queries, increasing evidence quantity dominates ordering, and capacity mainly reduces variance rather than flipping the sign of the effect.

Second, architectural design and model capacity affect this sensitivity but not uniformly. Larger models show lower variance (smoother \(\Delta\)F1) across both datasets, consistent with better context integration. However, they do not eliminate order effects in multi-hop settings. Regarding model architectures, we can see that across LLaMA-3.1, Mistral-Nemo, and Gemma-3, the mid-sized variants show larger positive \(\Delta\)F1 on HotpotQA at higher $k$ (reverse helps), whereas their larger models are less sensitive. On NQ, all three families are comparatively order-stable across k. These observations imply that ordering sensitivity arises from an interaction between dataset properties and how each architecture manages long-context reasoning, rather than from model size alone.

Finally, we analyse overall performance, setting aside the oracle model, which unsurprisingly achieves the highest scores but declines as more BM25 noise is introduced. We observe that the larger models, LLaMA-3.1:70B and Gemma-3:27B, consistently achieve the best F1 results across both datasets. Their performance remains stable across different topic sizes, indicating that increased model capacity contributes not only to improved accuracy but also to greater robustness against fluctuations in context size.

\section{Conclusions}

In this work, we revisit how context size and evidence ordering shape QA performance in RAG under a controlled, reproducible evaluation framework. We show that topic sampling is a major source of variance: conclusions about ordering and context size can shift when experiments are run on small topic subsets. Using HotpotQA and Natural Questions, we introduce a practical calibration procedure based on repeated subset sampling and fix topic sizes that yield stable trends at a feasible cost. Under this controlled setting, several patterns emerge. On single-hop NQ, performance generally improves as $k$ increases and is comparatively insensitive to order. On multi-hop HotpotQA, larger contexts help mainly when high-value evidence is placed in positions the model preferentially uses. Dense reranking also boosts small-$k$ performance and narrows ordering sensitivity. Larger models are typically more stable overall, but they do not eliminate ordering effects in multi-hop settings. Altogether, reliable evaluation is essential for interpreting order/size effects, and practitioners should prioritise retrieval quality and evidence placement over simply increasing context size. We release all code and configurations to support reproducible, order-aware RAG evaluation.

\section*{Computational resources}

Experiments were conducted using a private infrastructure, which has a carbon efficiency of 0.432 kgCO$_2$eq/kWh. A cumulative of 160 hours of computation was performed on hardware of type A100 SXM4 80 GB (TDP of 400W), generating an estimated 22.6 kgCO$_2$eq, with 0\% directly offset \cite{lacoste2019quantifying}.

\begin{acks}
All authors acknowledge funding from the Ministry of Science, Innovation and Universities of the Government of Spain (project PID2022-137061OB-C21, MCIN/AEI/10.13039/501100011033), as well as from the Department of Education, Science, Universities, and Vocational Training of the Xunta de Galicia (grant GRC ED431C 2025/49). CITIC, as a center accredited for excellence within the Galician University System and a member of the CIGUS Network, receives subsidies from the Department of Education, Science, Universities, and Vocational Training of the Xunta de Galicia. Additionally, it is co-financed by the EU through the FEDER Galicia 2021-27 operational program (Ref. ED431G 2023/01).
\end{acks}

\bibliographystyle{ACM-Reference-Format}
\balance
\bibliography{references}

\end{document}